\documentclass[prl,aps,twocolumn,showpacs,amsmath,amssymb]{revtex4}
\usepackage{graphicx}

\usepackage{dcolumn}
\usepackage{bm}
\usepackage{times}

\newcommand{\cs}{{\bf cs}}

\newcommand{\br}{{\bf r}}

\newcommand{\bk}{{\bf k}}

\newcommand{\df}{\stackrel{\rm def}{=}}

\newcommand{\kF}{k_{\scriptscriptstyle F}}

\newcommand{\Vsc}{V_{\rm sc}}

\newcommand{\Det}{\operatorname{Det}}

\newcommand{\smeq}{\! = \!}

\newcommand{\kf}{k_{\scriptscriptstyle F}}

\newcommand{\lambdaf}{\lambda_{\scriptscriptstyle F}}

\newcommand{\PSImagx}[2]{\includegraphics[width=#2]{#1}}

\newcommand{\Ssm}{{ \langle {\cal S} \rangle_{\Delta N}  }} 
\newcommand{\Si}{{ {\cal S}_i }} 

\newcommand{\varSi}{{{\rm Var}[{\cal S}_i]  }}

\begin{document}

\title{Residual Coulomb interaction fluctuations in chaotic systems: the boundary, random plane waves, and semiclassical theory} 
\author{Steven Tomsovic$^{1,}$\footnote{permanent address: Department
    of Physics and Astronomy, Washington State University, Pullman, WA
    99164-2814},  Denis Ullmo$^2$ and Arnd B\"acker$^3$}
 \affiliation{$^1$ Max-Planck-Institut f\"ur Physik komplexer Systeme,
   D-01187 Dresden, Germany} 
\affiliation{$^2$CNRS; Univ. Paris-Sud; LPTMS UMR 8626, 91405
  Orsay Cedex, France} 
\affiliation{$^3$Institut f\"ur Theoretische Physik, 
Technische Universit\"at Dresden,
01062 Dresden, Germany} 
\date{\today}

\begin{abstract}
  New fluctuation properties arise in problems where both spatial
  integration and energy summation are necessary ingredients.  The
  quintessential example is given by the short-range approximation to
  the first order ground state contribution of the residual Coulomb
  interaction.  The dominant features come from the region near the
  boundary where there is an interplay between Friedel oscillations
  and fluctuations in the eigenstates.  Quite naturally, the
  fluctuation scale is significantly enhanced for Neumann boundary
  conditions as compared to Dirichlet.  Elements missing from random plane wave modeling of chaotic eigenstates lead surprisingly to significant errors, which can be corrected within a purely semiclassical approach.  
\end{abstract}

\maketitle


The characterization of quantum systems with any of a variety of
underlying classical dynamics, ranging from diffusive to chaotic to regular, has
often demonstrated that the study of their statistical properties is
of primary importance.  Spectral fluctuations are a principle example
as they gave the first support to one of the main results linking
classical chaos and random matrix theory \cite{MehtaBook04}, the
Bohigas-Giannoni-Schmit conjecture \cite{Bohigas84,Bohigas91}.
Needless to say, the statistical properties of eigenfunctions are also
a subject of paramount interest \cite{Berry77a, Voros79,
  McDonald79, McDonaldThesis, Heller84, Heller91, Srednicki96,
  Hortikar98prl, Mirlin00PhysRep}.

For chaotic systems, a widely accepted starting point for the treatment
of eigenfunction fluctuations {\em locally}, such as the
amplitude distribution or the two-point correlation function $c(|\br -
\br'|) = \langle \psi(\br) \psi(\br') \rangle$ of a given
eigenfunction $\psi$, is a modeling in terms of a random superposition
of plane waves (RPW) \cite{Berry77a,Voros79}.  For two-degree-of-freedom
systems, $c(r)$ can be
understood as being given approximately by a Bessel function.  For
distances $|\br - \br'|$ short compared to the system size, and in the
absence of effects related to classical dynamics
\cite{Heller84,Heller91,Bohigas93}, this is roughly observed in
numerical \cite{McDonaldThesis,McDonald88,Baecker02} and
experimental \cite{KimBarKuhSto2003} studies.  Our interest
here is in statistical properties of eigenfunctions going beyond local quantities such as $c(r)$.

One motivation for the introduction of these new statistical measures
is to study the interplay between interferences and
interactions in mesoscopic systems.  
For typical electronic densities, the
screening length is close to the Fermi wavelength
$\lambdaf$, and the screened Coulomb interaction can be approximated
by the short range expression
$
\Vsc(\br-\br') = ({2\nu})^{-1} {F_0^a} \delta(\br-\br') 
$
with $2\nu$ the mean local density of states, including spin degeneracy, 
($\nu \smeq m/2\pi\hbar^2$ for $d \smeq 2$) 
and $F_0^a$ the dimensionless Fermi liquid parameter
\cite{PinesNozieresVol1}, a constant of order one.    

To this level of approximation, the first order ground
state energy contribution of the residual interactions can be expressed in the form
$
\delta E^{\rm RI} = (2\nu)^{-1} {F_0^a } \int d\br \, N_\uparrow(\br)
N_\downarrow(\br),
$
with $N_\sigma$ the unperturbed ground state density of particles with
spin $\sigma$. From this expression, it is seen that the increase of interaction energy associated with the
addition of  an extra electron is related to 
\begin{equation}
\label{eq:sinr}
\Si = {\cal A}\int {\rm d}\br\
\left|\Psi_i(\br) \right|^2N(\br;E_i^+) \; ,
\end{equation}
with $
N(\br ; E) \equiv 
\sum_{i=1}^\infty \left|\Psi_i(\br)\right|^2 
\theta \left( E - E_i \right) $ 
and the understanding that $E_i < E_i^+ < E_{i+1}$.  Our goal in this
letter is to study the fluctuation properties of the $\Si$,
concentrating on the case of two dimensional
billiards with either Dirichlet or Neumann boundary conditions.   For them, $\cal A$ is the billiard area.

The dominant contributions to ${\cal S}_i $
(and its fluctuations) originate from the Friedel oscillations of the
density of particles near the boundary.  For billiards systems, they
can be expressed as $ N_{\rm Friedel}(\br;E) = \frac{N_W(E)}{{\cal
    A}}\left[ 1 \pm \frac{J_1(2kx)}{kx} \right] $~\cite{Baecker98}, with $x$ the
distance from the boundary, the $+$ and $-$ sign corresponding
respectively to Neumann and Dirichlet boundary
conditions, and $N_W(E)$ refers to
leading term of the Weyl formula, $N_W(E) = \nu{\cal A}  E$.  To leading order we can therefore use the
approximation
\begin{equation} \label{eq:Si}
{\cal S}_i = i \pm i \int {\rm d}\br \ \left[ 
  \frac{J_1(2 \kf x)}{\kf x}  |\Psi_i(\br)|^2 \right] \;.
\end{equation}
To proceed, a description of
the fluctuations of $\left|\Psi_i(\br) \right|^2$ is also required. These are obtained ahead
using a semiclassical approach closely related to the
Gutzwiller trace formula.   However, it is useful first to
consider the oft-employed RPW description, which very interestingly turns out to lack a couple of crucial ingredients.  Nevertheless, it sheds  light on the mechanism governing
the fluctuations under study.
 
Within RPW~\cite{Berry77a,Voros79} 
eigenstates are represented, in the absence of any symmetry, by a
random superposition of plane waves $\sum_l a_l \exp(i \bk_l
\cdot\br)$ with wave-vectors of fixed modulus $|\bk_l|=\kf$
distributed isotropically.  Time reversal invariance introduces a
correlation between time reversed plane waves such that the
eigenfunctions are real.  Similarly, the presence of a planar boundary
imposes a constraint between the coefficients of plane waves related
by a sign change of the normal component of the wave-vector $\bk_l$
\cite{Berry02,Urbina04}.
Near a boundary, and using a system of coordinates $\br = \hat {\bf x}
+ \hat {\bf y}$ with ($\hat {\bf x}$,$\hat {\bf y}$) the vectors
respectively parallel and perpendicular to the boundary, eigenfunctions are mimicked
statistically by a superposition,
\begin{equation}
\label{eq:rpwm}
\psi_i(\br) = \frac{1}{N_{\rm eff} }\sum_{l=1}^{N_{\rm eff}}
a_l \cs \left( 
  {\bf k}_l \cdot \hat {\bf x} \right) 
   \cos\left( {\bf k}_l\cdot \hat{\bf y} +
  \varphi_l\right) 
\end{equation}
where $\cs(\cdot) \df \sin(\cdot)$ for Dirichlet and $\cos(\cdot)$ for
Neumann boundary conditions.  The phase angle
$\varphi_l$, the orientation of the wave vector ${\bf k}_l$, and the real amplitude $a_l$ with $\langle a_l a_{l'} \rangle = \delta_{l {l'}} \sigma^2$ are all chosen randomly.  Normalization of the wave-functions fixes the variance $\sigma^2$ through the relation
\begin{equation}
\label{eq:normalization}
1= \int d\br \, \left<\left|\psi_i(\br)\right|^2\right> 
= \frac{ { \cal A} \sigma^2}{ 4N_{\rm eff} }\left( 1 \pm \frac{{\cal L} }{
    2\kf {\cal A}} \right) \; ,
\end{equation}
where $\cal L$ is the perimeter.

The variance is a natural measure of the fluctuations.  To leading order
\begin{equation*}
\begin{split}
\label{eq:var}
{\rm Var} & \left({\cal S}_i  \right) =   \frac{ i^2}{ {\cal A}^2} \int {\rm
  d}\br_1  {\rm d}\br_2\  \frac{J_1(2 \kf x_1) }{\kf x_1}
\frac{J_1(2\kf x_2)
  }{\kf x_2} \\
 \times & \left[ \left<|\Psi_i(\br_1)|^2 |\Psi_i(\br_2)|^2 \right> 
  -\left<|\Psi_i(\br_1)|^2\right>\left<
    |\Psi_i(\br_2)|^2\right>\right] \; .
\end{split}
\end{equation*}
The fluctuations are thus given by pair-wise
correlating the random plane wave coefficients $a_n$  such that
$
\langle a_l a_{l'} a_m a_{m'} \rangle = \langle a_l a_{l'} \rangle
\langle a_m a_{m'} \rangle + \langle a_l a_m \rangle \langle a_{l'}
a_{m'} \rangle + \langle a_l a_{m'} \rangle \langle a_{l'} a_m \rangle
= \sigma^4\left(\delta_{ll'} \delta_{mm'} + \delta_{lm}
  \delta_{l'm'} + \delta_{lm'} \delta_{l'm} \right) $.  One obtains in
this way
\begin{equation}
\label{eq:varianceexp}
\begin{split}
{\rm Var} & \left[{\cal S}_i \right] =  \frac{8 i^2}{{\cal A}^2}
\frac{1}{N_{\rm eff}^2} \sum^{N_{\rm eff}}_{l , m=1} \int d\br_1 d\br_2
\frac{J_1(2\kf x_1)}{\kf x_1} \frac{J_1(2\kf x_2)}{\kf x_2} \\
& \cs({\bf k}_l\cdot {\bf x}_1) \cs({\bf k}_l\cdot {\bf x}_2)  
\cs({\bf k}_m\cdot {\bf x}_1) \cs({\bf k}_m\cdot {\bf   x}_2) \\
& \cos\left[ {\bf k}_l \cdot ({\bf y}_1- {\bf y}_2)\right]
\cos\left[ {\bf k}_m \cdot ({\bf y}_1- {\bf y}_2)\right] \; .
\end{split}
\end{equation}
Performing the integral gives
\begin{equation} {\rm Var}[{\cal S}_i]  = 
\frac{\kf {\cal L} }{4\pi^3} \langle \lambda^2(\theta) \rangle_\theta
\label{eq:VarRPW} 
\end{equation}
where we have introduced the function 
$ \lambda (\theta) \df \left[{1 \pm
      |\sin(\bar\theta)|}\right]/|\cos(\bar\theta)| $
 and the expression for 
$\langle \lambda^2(\theta) \rangle_\theta \equiv \int_0^1 d(\sin \theta)
\lambda^2(\theta)$ is given ahead in Eq.~(\ref{lambdatheta}).

\begin{figure}[t]
\PSImagx{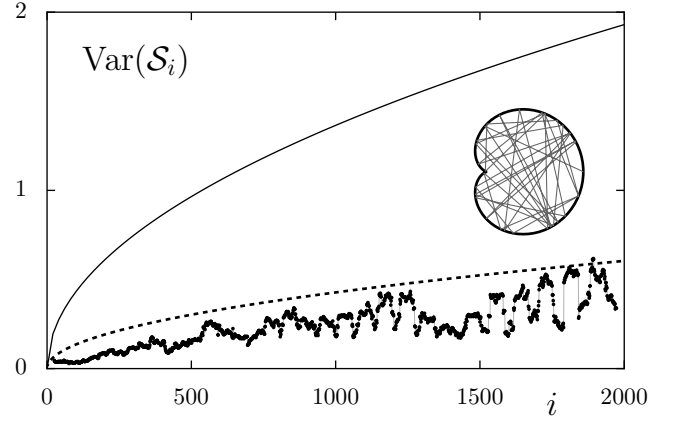}{8.5 cm}
\caption{Variance of ${\cal S}_i$ for the cardioid billiard with
  Dirichlet boundary conditions.  The solid line is the RPW
  expression, Eq.~(\ref{eq:VarRPW}), the dashed line is the
  semiclassical prediction, Eq.~(\ref{eq:Varsc}), and the discrete
  points are for the cardioid billiard (shown in inset).}
\label{fig:varianceSi}
\end{figure}

In Fig.~\ref{fig:varianceSi}, the variance of $\Si$ as a function of
$i$ is represented for a chaotic system, the cardioid billiard with
Dirichlet boundary 
conditions~\cite{Robnik84,Baecker95}.  In this case, quite surprisingly given the history of modeling chaotic eigenstates within the RPW framework, its predictions significantly overestimate the fluctuations.  In fact, two important elements are missing from this approach, both of which are addressed properly within a purely semiclassical approach.

One difficulty immediately encountered with a semiclassical treatment
of the  $\Si$ is that their computation implies addressing the
fluctuations of  individual wave-functions, whereas the semiclassical
approximations valid for chaotic systems of use here  
converge only for (locally) smoothed  quantities.  This difficulty may
be overcome by following the spirit of Bogomolny's
calculation~\cite{Bogomolny88}, and introducing a local energy
averaging, 
\begin{equation}
\label{eq:smean}
\langle {\cal S} \rangle_{\Delta N} \df \frac{1}{\Delta N} \sum_{E
  -\frac{\Delta E}{2} < E_i < E +   \frac{\Delta E}{2}}   \Si \;  ,
\end{equation}
with $\Delta N \df   N(E \!+ \! \frac{\Delta E}{2})\!  - \!  N(E\! -\! \frac{\Delta  E}{2})$.  This generates the relation
\begin{eqnarray} 
\label{eq:transmuttation}
&& {\rm Var} [\langle {\cal S} \rangle_{\Delta N}] \df 
 \overline{\left (\Ssm  - \overline{\cal S}\right)^2} \nonumber \\
&=& \frac{1}{\Delta N}\varSi + \left(1- \frac{1}{\Delta N}\right) {\rm Covar}\left[{\cal S}_i {\cal S}_{i+j}\right] .
\end{eqnarray}
assuming that there is translational invariance (smooth, uniform behavior) locally in indices ($i,j$).  See~\cite{Ullmo08} for further discussion for the relevance of this assumption.  Computing the smoothed quantity $\Ssm$, for which 
semiclassical approximations are convergent, one can therefore extract
the variance of the $\Si$ from the scaling in
$\Delta N$ of ${\rm Var} [\langle {\cal S} \rangle_{\Delta N}]$.

Our starting point for this calculation is 
\begin{eqnarray}
\label{eq:wf1} 
{\cal A}  \langle |\Psi_i(\br)|^2 \rangle_{\Delta N}
&=&  1  \pm J_0(2\kf  x) -   \frac{1}{\nu } \frac{1}{\pi} {\rm Im} \langle \tilde G^{\rm 
  osc} (\br,\br,E) \rangle_{\Delta E} \nonumber  \\  
&&  -  \frac{1 \pm J_0(2\kf  x)}{{\cal A} \nu} \langle \rho^{\rm osc}\rangle_{\Delta E} \; ,
\end{eqnarray} 
valid near the boundary, which differs from the expression given by Bogomolny
\cite{Bogomolny88} only through the inclusion of the  Bessel
function $J_0(2\kf x)$ accounting for the Friedel oscillations. Here, the energy smoothing indicated is similar to above except normalized by the energy range, and 
$\rho^{\rm osc}$ is the oscillating part of the density of states,
given semiclassically as a sum over periodic orbits
\begin{equation} 
\label{eq:rho_osc} 
\rho^{\rm osc}(E) =  \frac{1}{\pi\hbar}  \sum_{\gamma={\rm
    p.\, o.}}\tau_\gamma \frac{ \cos\left(
 \frac{S_\gamma(E^+_i)}{\hbar} - \sigma_\gamma \frac{\pi}{ 2}\right) }{ 
 \left| \Det\left(M_\gamma - {\bf 1}\right)\right|^{1/2}} \; .
\end{equation}
Similarly, the diagonal part of the Green's function $G^R(\br,\br,E)$
is expressed as a sum over closed (not necessarily periodic) orbits
\begin{equation} 
\label{eq:G1/2classic}
\begin{split}
G^{\rm  osc}(\br,\br,E) & \simeq 
\frac{1}{i\hbar}   \frac{1}{\sqrt{2i\pi\hbar}  }
\sum_{\mu: \br \to \br}  \frac{1}{ \sqrt{|\dot x \dot x'
    m_{12,\mu}|}} \\ & \times \exp\left[ 
  \frac{i}{\hbar} S_\mu(\br,\br,E) - i \frac{\pi}{2} \eta_\mu\right ]
\; .
\end{split}
\end{equation}
In the above expressions, $S$ is the action integral along the orbit,
$\tau$ the period, $m_{12} \equiv \partial y^{\rm final}/\partial
p_y^{\rm initial}$, $M_\gamma$ the monodromy matrix, and
$\sigma_\gamma$, $\eta_\mu$ are Maslov indices.  The tilde on the
Green's function in Eq.~(\ref{eq:wf1}) furthermore indicates that the
short orbit giving rise to Friedel oscillations, namely the one
bouncing off the boundary and returning directly to its initial
location, is excluded from the semiclassical sum (as in
Eq.~(\ref{eq:wf1}) it is already taken into
account by the Bessel function).

\begin{figure}[t]
\PSImagx{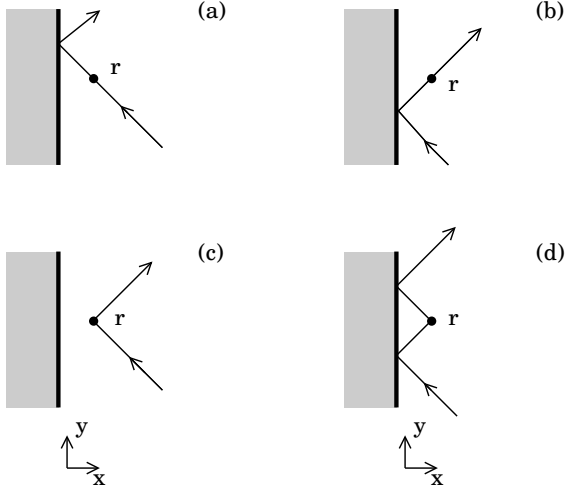}{7.5cm}
\caption{ In the semiclassical
  sums, the four orbits which, as $x \to 0$, coalesce into the same
  [nearly] periodic orbit, need to be aggregated. The top row
  correspond to two (nearly) periodic orbits. The bottom row
  correspond to two non periodic orbits ($p'_x =- p_x)$.}
\label{fig:orb_clust}
\end{figure}

Inserting Eq.~(\ref{eq:wf1}) into Eq.~(\ref{eq:Si}), and performing
the integral over space, it is important to note that the range of
integration in the direction perpendicular to the boundary is short,
even on the quantum scale, and that therefore a stationary phase
condition should be imposed only on the parallel direction.  Gathering
however the four orbits shown in Fig~\ref{fig:orb_clust}, and
labelling their total contribution with the index $\gamma$ of the periodic
orbit to which they converge as $x \to 0$, one obtains, following the usual
steps of the derivation of the Gutzwiller trace formula,
\begin{eqnarray} 
\label{eq:semiresult}
\langle {\cal S} \rangle_{\Delta N}  - \overline{\cal S}  && =  \pm
\frac{1}{\pi}  
\sum_{\gamma = {\rm p.\, o.}}
\frac{\cos\left[\frac{ S_\gamma(E)}{\hbar} -  \bar\nu_\gamma
      \frac{\pi}{ 2}\right]}{\sqrt{|\Det(M_\gamma-1)|}}  \times \nonumber \\
&&   {\rm  sinc}\left(\frac{\tau_\gamma   \Delta E}{2 \hbar} \right)  
\sum_{ l \in   \gamma}    \left(
  \lambda(\theta_l) - \tilde \lambda_\gamma  \right)  ,
\end{eqnarray}
where $\lambda(\theta_l)$ is the function introduced in
Eq.~(\ref{eq:VarRPW}) evaluated at the angle at which the periodic
orbit $\gamma$ strikes the boundary on the $l^{th}$ bounce, and
$\tilde \lambda_\gamma \df \frac{{\cal L} 
  \ell_\gamma }{ 2{\cal A} n_\gamma}\left( 1\pm \frac{2}{\pi}\right) $, with
  $\ell_\gamma$ the total length and $n_\gamma$ the number of bounces
  of the orbit $\gamma$.

Eq.~(\ref{eq:semiresult}) can be used directly to compute the
  Fourier transform of the $\Si$, yielding a structure very similar to
  that of the density of states Eq.~(\ref{eq:rho_osc}).  Both
  Fourier transforms can be done so as to have peaks with the same shape and
  positions, but different amplitudes.  Eq.~(\ref{eq:semiresult}) can
  be furthermore used to compute the variance of $\langle {\cal S}
  \rangle_{\Delta N}$.  To begin, square Eq.~(\ref{eq:rho_osc}), use the
  diagonal approximation in which orbits are paired only with
  themselves or their time reverse symmetric partner, and apply the Hannay
  Ozorio de Almeida sum rule \cite{OzorioBook}
\begin{equation*}
\sum_{\gamma = {\rm p. o.}} \frac{n} {|\Det(M_l-1)|} = 1 \; ,
\end{equation*}
where only periodic orbits of $n$ bounces from the boundary are
included.  Then for the long orbits relevant here, replace the mean
length $\ell_\gamma/n_\gamma$ between two successive bounces for a
specific orbit $\gamma$ by $\bar d = {\pi {\cal A}}/{\cal L} $, the
mean length between bounces averaged for all initial conditions on the
boundary.  Noting $\left\langle \lambda(\theta) - \tilde \lambda
\right\rangle=0$, this gives
\begin{equation}
\label{eq:VarSDN}
\overline{\left( \langle {\cal S} \rangle_{\Delta N} - \overline{\cal
      S}  \right)^2} 
  =  
\frac{\kf {\cal L} }{2 \pi^3} 
\left\langle \left(\lambda(\theta) -  \tilde \lambda \right)^2 \right
\rangle_\theta \, 
\frac{1}{\Delta N}  \; .
\end{equation}
As before, angle average $\langle \cdot
\rangle_\theta$ is over the measure $d\sin\theta$.

The structure of this result is quite interesting.  The surviving
contribution comes from what may be called the ``diagonal-diagonal''
terms, i.e.\ pairing not only of the same orbit but of the same bounce
from the boundary.  This term contributes only to the variance as it
scales as $1/\Delta N$; see Eq.~(\ref{eq:transmuttation}).  The terms
that would be called ``diagonal-off-diagonal'' give a vanishing
contribution.  In addition, note the absence of a constant term.  It
implies that the covariance vanishes, which is consistent with our
cardioid billiard calculations (not shown here).

Thus, from Eq.~(\ref{eq:transmuttation}) and the above considerations
\begin{equation}
\mbox{Var}({\cal S}_i) = \frac{\kf {\cal L} }{ 2 \pi^3} \times 
\left\langle \left(\lambda(\theta) -  \langle
  \lambda\right\rangle_\theta  \right)^2 \rangle_\theta 
\label{eq:Varsc}
\end{equation}
where 
 \begin{equation}
 \label{lambdatheta}
\begin{split}
 & \left\langle \left(\lambda(\theta) -  \langle
   \lambda\right\rangle_\theta  \right)^2 \rangle_\theta \\
 & = \left\{
   \begin{array}{ll} 
 (2 \ln 2 -1) -  \left(\frac{\pi}{2} -1 \right)^2 \simeq 0.06 & \mbox{Dirichlet } \\
 (2 \ln 2 -1) -  \left(\frac{\pi}{2} -1 \right)^2 + 4\left(\ln
   \frac{\pi \kf {\cal A} }{ 2{\cal L}} -\frac{\pi}{2}\right) \quad &
 \mbox{Neumann}  
 \end{array}
 \right. 
 \end{split}
 \end{equation}
 Comparing this expression with Eq.~(\ref{eq:VarRPW}), we see that the
 semiclassical approach and the random plane wave model lead to the
 same result, except for two differences.  First, a factor two in the
 prefactor can be traced to dynamical correlations missed by RPW.  The
 nature of these correlations, which are somewhat subtle, will be
 discussed in \cite{Ullmo08}.  Second, the mean square $\langle
 \lambda^2(\theta) \rangle_\theta$ has been replaced by the variance
 of $\lambda(\theta)$, giving now a much better agreement with the
 numerically evaluated ${\cal S}_i$ (see Fig.~\ref{fig:varianceSi}).

In Eq.~(\ref{eq:semiresult}), the term
 proportional to $\lambda(\theta_\gamma)$ can be seen to arise from
 $G^{\rm osc}$ in Eq.~(\ref{eq:wf1}), while the one proportional to
 $\tilde \lambda_\gamma$ originates from $\rho^{\rm osc}$. The random
 plane wave result is thus, in some sense, ignoring the latter
 contribution.  As $\rho^{\rm osc}$ has no spatial 
 dependence, its role is not to describe the variations of the
 wave-function, but rather to ensure their normalization (as can be
 seen readily by integrating Eq.~(\ref{eq:wf1}) over space).
 Therefore the main reason of the failure of the RPW approach is due to the
 lack of individually normalized wavefunctions, and more precisely to
 the fact that Eq.~(\ref{eq:normalization}) imposes the normalization
 of the wavefunctions only on average \cite{Urbina07}.  

Equation~(\ref{eq:Varsc}) emphasizes the importance of the
boundary conditions.  Indeed, the variance of $\lambda(\theta)$ is
extremely small for Dirichlet boundary conditions, which has to be
expected since the wave-functions are zero near the
boundary.  As a consequence, and as seen in Fig.~\ref{fig:varianceSi},
the fluctuations remain smaller than one even for a relatively large
number of particles, in spite of the linear $(\kF {\cal L})$
dependence of the variance.  On the other hand, Neumann boundary conditions yield a logarithmic divergence which can be considered in practice as a constant somewhat larger than one.
Fluctuations in this case are greatly enhanced with respect to
the Dirichlet case.

To conclude, ``integrated'' wavefunction statistics are introduced, in
part motivated by the need to understand the effect of interactions on
ground state properties in quantum dots.  A significant part of their
fluctuation properties can be understood with basic RPW modeling.
However, a semiclassical framework is developed here, which indicates
missing ingredients of RPW, and in particular that lack of individual
state normalization leads to a significant overestimate of the
fluctuations.  It turns out furthermore that boundary conditions,
which are often not discussed in the context of mesoscopic systems,
play an important role.  

We stress finally that only the extreme limit of strongly chaotic
systems has been treated here.  For less developed chaos, or systems
with some regular dynamics, where some eigenstate localization exists,
there may be significant enhancements in the fluctuations and
correspondingly greater effects on ground state
properties~\cite{Ullmo03}.  The consequences for experimentally
realizable systems and for such systems with some form of eigenstate
localization is left for future study.

One of us (ST) gratefully acknowledges support from US National
Science Foundation grant  PHY-0555301. 

\bibliography{rmt,nano,general_ref,quantumchaos,furtherones,classicalchaos}

\end{document}